\documentclass[onecolumn,preprint,showpacs,showkeys,superscriptaddress]{revtex4-1}
\usepackage{amssymb}
\usepackage{amsfonts}
\usepackage{amsmath}
\usepackage{graphicx}
\usepackage{epstopdf}
\usepackage{float}
\usepackage{relsize}

\begin{document}

\title{Electrostatic self-interaction of charged particles in the space-time of a cosmic string in the context of gravity's rainbow }
\author{L. C. N. Santos}
\email{luis.santos@ufsc.br}
\affiliation{Departamento de F\'isica, CCEN - Universidade Federal da Para\'iba; C.P. 5008, CEP  58.051-970, Jo\~ao Pessoa, PB, Brasil}
\author{V. B. Bezerra}
\email{valdir@fisica.ufpb.br}
\affiliation{Departamento de F\'isica, CCEN - Universidade Federal da Para\'iba; C.P. 5008, CEP  58.051-970, Jo\~ao Pessoa, PB, Brasil}

\begin{abstract}
We analyze the electrostatic self-energy of a point like electrically charged particle induced by a cosmic string in the context of gravity's rainbow, as well the electrostatic self-force on this particle. The possibility of the solution associated with a charged particle to be altered by modifications in  dispersion relations of the space-time is discussed. We show that the self-energy depends on the rainbow functions and that this quantity can either increase or decrease depending on the rainbow function chosen, as compared with analogous result in the framework of general relativity. With respect to the self-force, its dependence with the rainbow functions is also pointed out.
\end{abstract}

\maketitle

\preprint{}

\section{introduction}
\label{sec1}
It is known that when a charged particle is placed at rest in a gravitational field appears an electrostatic self-energy and self-force on the localization of the charge \citep{rainbow18,rainbow16,rainbow20,rainbow21}. These phenomena of self-energy are very interesting and has attracted a lot of attention (for a review see \cite{self1}). When we consider the Minkowski space-time, the origin of the self-force is connected with inertial properties of the electrostatic field. In the framework of general relativity, the electrostatic as well as the gravitational self-forces also appear. In this case, there is an additional difficult due to the fact that each kind of energies create a gravitational field, including the one associated with the electromagnetic field. 

In order to understand the origin of such self-interaction in a designed space-time, we observe that the electromagnetic field of a particle in a gravitational background polarizes the "medium" and as a consequence a self-force appears \cite{rainbow16}. In general relativity, although a straight cosmic string \cite{string3,rainbow23} produces no local gravitational force on a surrounding charged particle, there is an interaction between the cosmic string and the particle placed at rest in this space-time which is of global origin. It is believed that this interaction results from the fact that the field of a charged particle, in this space-time, is deformed due to the global features of the cosmic string space-time and as a consequence appears a self-force which depends on the angular deficit of space-time geometry which has topology cone$\times\mathbb{R}^2$ \cite{rainbow17}. The case of the electrostatic self-force on a point charge at rest
in an arbitrary space-time with cylindrical symmetry in the linear approximation in the Newtonian constant has been studied in Ref. \cite{rainbow19} where the authors obtained a self-interaction potential energy in terms of the components of the metric. In the particular case in which the space-time considered is generated by an infinite, static and cylindrically symmetric cosmic string, the electrostatic self-force was obtained by Linet \cite{rainbow15}.

In recent years, modified theories of gravity have been proposed in order to construct a formalism compatible with a quantum treatment of gravity, among other proposals. In this sense, the gravity's rainbow \cite{rainbow1} is interesting because it incorporates modifications in dispersion relations that allow to explain, in principle, threshold anomalies in ultra high cosmic rays and TeV photons \cite{rainbow11,rainbow2}. Effects of the gravity's rainbow on various physical systems were studied recently as, for example, the quantum vacuum fluctuations of the scalar field, the renormalized vacuum energy in the Friedmann-Robertson-Walker space-time \cite{rainbow8}, the geodesic structure of the Schwarzschild black hole  \cite{rainbow7} and the hydrostatic equilibrium equation of stars \cite{rainbow6}.  Other possibility to use this gravitational theory is to consider a relevant aspect not yet explored in this context which consists in the determination of the electrostatic self-energy and self-force that appears when a charged particle is placed in the vicinity of a cosmic string in the framework of gravity's rainbow \cite{rainbow12}. In this case, as the cosmic string solution has the same global features of the corresponding solution in the context of the general relativity, it is expected that the self-interaction has the same global origin as in the last case. Additionally, the intensity if the self-force should depend strongly on the gravity’s rainbow functions chosen as is pointed out in this paper. In fact, depending on the rainbow function, the results represent a lower or upper bound for the intensity of the self-force as compared with similar results in the context of the general relativity.

The organization of this paper is as follows. In Sec. \ref{sec2}, we present the  Maxwell's equations in curved space-time and obtain the electrostatic self potential. In Sec. \ref{sec3} we study the electrostatic self-energy and self-force and obtain a expression corresponding to these quantities for a particle at rest in the cosmic string space-time in the context of gravity's rainbow. Finally, in Sec. \ref{sec4}, we discuss the obtained results. 
\section{Electrostatic self-potential}
\label{sec2}
 Let us start presenting briefly the form of the equations associated to the electromagnetic field in a arbitrary space-time background. Applying the principle of general covariance to Maxwell's equations in the Minkowski space-time it is possible to obtain the equations that describe the behavior of the electromagnetic field in an arbitrary geometry. The resulting equations in general a curved space-time reads as
 \begin{equation}
 \frac{1}{\sqrt{-g}}\partial_{\nu}(F^{\mu\nu}\sqrt{-g})=4\pi j^{\mu},
 \label{eq1}
\end{equation}  

\begin{equation}
F_{\mu\nu}=\partial_{\mu}A_{\nu}-\partial_{\nu}A_{\mu},
\label{eq2}
\end{equation}
where $j^{\mu}$ and $A^{\mu}$ are the four-vector current and the electromagnetic four-vector potential, respectively. In the case of a static charge in a curved background, we have the relations $j^{k}=0$, $A^{k}=0$ and $j^{0}=q\frac{\delta(\vec{x}-\vec{x}_{0})}{\sqrt{-g}}$. 

The space-time metric of a cosmic string in the gravity's rainbow context is given by \cite{rainbow12}
\begin{equation}
ds^2=f_{1}(E/E_{p})^{-2}dt^2  - f_{2}(E/E_{p})^{-2}d\rho^{2} - f_{2}(E/E_{p})^{-2}\alpha^{2}\rho^{2}d\phi^{2} -f_{2}(E/E_{p}) ^{-2}dz^{2},
\label{eq3}
\end{equation}
where $\alpha=1-4\mu$ is a parameter associated to the deficit angle of the cosmic string , where $\mu$ is the linear mass density of the cosmic string and $f_{1}(E/E_{p})$  and $f_{2}(E/E_{p})$ are the rainbow functions, with $E/E_{p}$ being the ratio between the energy of the test particle and the Planck energy. Here we are interested in two most adopted rainbow functions. The first one is giving by 
\begin{equation}
f_{1}(E/E_{p})=1, \qquad f_{2}(E/E_{p}) =  \sqrt{1+(E/E_{p})^{2}}.
\label{eq4}
\end{equation}
This rainbow function was considered to investigate the formulation of  doubly-special
relativistic theories  in which the speed of photons acquires a dependence on energy or wavelength \cite{rainbow24}. As a second set of rainbow function, we will consider the following
\begin{equation}
f_{1}(E/E_{p}) = \frac{e^{E/E_{p}}-1}{E/E_{p}}, \qquad f_{2}(E/E_{p}) = 1. 
\label{eq5}
\end{equation}
The exponential behavior of this function is applied to study of the problem of gamma-ray burst phenomena in the universe \cite{rainbow13,rainbow14}. 

Now, let us write the Maxwell's equations given by Eq. (\ref{eq1}) in the space-time described by the metric (\ref{eq3}) in order to obtain, for the point charge $q$, located at $\rho=\rho_{0}, \phi=\pi, z=0$, the equation for the electrostatic potential $V(\rho,\phi,z)$,  which can be written as
\begin{equation}
\left(\frac{\partial^{2}}{\partial \rho^{2}} + \frac{1}{\rho}\frac{\partial}{\partial\rho} + \frac{1}{\alpha^{2}\rho^{2}}\frac{\partial^{2}}{\partial\phi^{2}} + \frac{\partial^{2}}{\partial z^{2}}\right)V(\rho,\phi,z) = -\frac{4\pi qf_{2}}{\alpha\rho f_{1}}\delta(\rho-\rho_{0})\delta(\phi-\pi)\delta(z). 
\label{eq6}
\end{equation}
This nonhomogeneous differential equation permits us to obtain the electrostatic potential associated to a charged particle with a modified dispersion relation  due to  gravity's rainbow. Using the definitions
\begin{equation}
\theta = \alpha\phi, \qquad  \epsilon=\frac{ f_{1}}{4\pi f_{2}}, 
\label{eq6b}
\end{equation}
Eq. (\ref{eq6}) reduces to the usual potential equation in the Minkowski space-time where the dispersion relations are not deformed. Similarly to the case of a charged particle in the space-time of a cosmic string \cite{rainbow15} in general relativity, the electrostatic potential in the present case, should satisfy the same boundary conditions as the ones given in Eq. (6) of Ref. \cite{rainbow15}. Thus, we have
\begin{align}
V(\rho,0,z) & = V(\rho,2\pi\alpha,z), \label{eq7}\\
\frac{\partial V(\rho,0,z) }{\partial\theta} &=\frac{\partial V(\rho,2\pi\alpha,z)}{\partial\theta}=0. \label{eq8}
\end{align} 
A solution that satisfies these boundary conditions is given in Ref. \cite{rainbow15} In the context of general relativity. Thus, generalizing the result of \cite{rainbow15} to the present case, the  potential function can be written in an integral form as
\begin{align}
V(\rho,\phi,z) & =  \frac{f_{2}q}{f_{1}(2\pi\alpha)\sqrt{2\rho \rho_{0}}}\int_{x_{0}}^{\infty} \left[\frac{\sinh (x/2\alpha)}{\cosh (x/2\alpha)+\sin (\theta /2\alpha)} \right. +\nonumber\\
& \left.  \frac{\sinh (x/2\alpha)}{\cosh (x/2\alpha) - \sin (\theta /2\alpha)}\right] \frac{dx}{(\cosh x-\cosh x_{0})^{1/2}},
\label{eq9}
\end{align}
where the variable $x_{0}$ is defined by $\cosh x_{0} =  (\rho^2+\rho_{0}^{2}+z^{2})/2\rho\rho_{0}$.  Note that if we compare Eq. (\ref{eq9}) with similar result given by Eq. (7) of Ref. \cite{rainbow15}, we may conclude that the only modification generated by the gravity's rainbow is to introduce an effective charge $q_{eff}\equiv \frac{f_{2}}{f_{1}}q$. Thus, all the results follows straightforwardly. We may go back to the original  variables in Eq. (\ref{eq9}) by using definitions of $\theta$ and $\epsilon$ in Eq. (\ref{eq6b}), which results in
\begin{equation}
V(\rho,\phi,z)=\frac{ f_{2}q}{\pi\alpha f_{1}\sqrt{2\rho\rho_{0}}}\int_{x_{0}}^{\infty}\frac{\sinh (x/\alpha)}{[\cosh (x/\alpha)+\cos (\phi)]}\frac{dx}{(\cosh x-\cosh x_{0})^{1/2}}.
\label{eq10}
\end{equation}
This is the general solution that may be written as the combination of a regular solution $S$, and an irregular solution,  $V_{0}$. Mathematically, it can be written as
\begin{equation}
V(\rho,\phi,z) = V_{0}(\rho,\phi,z) + S(\rho,\phi,z).
\label{eq11}
\end{equation}
As in Ref. \cite{rainbow15}, the function $V_{0}(\rho,\phi,z)$, in the neighborhood of the charge, corresponds to the solution associated with the usual Coulomb potential in the Minkowski space-time. Following straightforwardly the procedure adopted by Linet \cite{rainbow15} to determine the function $S(\rho,\phi,z)$, let us write $V_{0}(\rho,\phi,z)$ in an integral form 
\begin{align}
V_{0}=\frac{ f_{2}q}{\pi f_{1}\sqrt{2\rho\rho_{0}}}&\int_{x_{0}}^{\infty}\frac{\sinh (x)}{[\cosh (x)+\sin (\alpha\phi)\sin (\alpha\pi) - \cos (\alpha\phi)\cos (\alpha\pi)]}\times \nonumber\\
&  \frac{dx}{(\cosh x-\cosh x_{0})^{1/2}},
\label{eq12}
\end{align}
which is more appropriate to set the expression for $S(\rho,\phi,z)$ from Eq. (\ref{eq11}), by using Eqs. (\ref{eq10}) and (\ref{eq12}).
\section{ Electrostatic self-energy}
\label{sec3}
The potential term $S(\rho,\phi,z)$ is nonzero for the general case in which $\alpha\neq 1$. Particularly, in the case of a cosmic string space-time, $\alpha$ should be in the interval $0<\alpha<1$. The case where $\alpha=1$ corresponds to the Minkowski space-time case in which we must recover the appropriate results. From Eq. (\ref{eq11}) and using the definition of the electrostatic energy $U=\frac{1}{2}qS(\rho_{0},\pi,0)$, we find that
\begin{equation}
U=\frac{q}{2}(V-V_{0})=\frac{q^{2}}{4\pi\rho_{0}}\Lambda(\alpha,E/E_{p}),
\label{eq13}
\end{equation}
where $\Lambda$ depend on the parameter $\alpha,$ as well as on the rainbow's functions $f_{1}(E/E_{p})$ and $f_{2}(E/E_{p})$, and is defined by the equation
\begin{equation}
\Lambda(\alpha,E/E_{p}) = \frac{f_{2}(E/E_{p})}{f_{1}(E/E_{p})}\int_{0}^{\infty}\left[ \frac{\sinh (x/\alpha)}{\alpha [\cosh (x/\alpha)-1]} - \frac{\sinh (x)}{\cosh (x) - 1}\right]\frac{dx}{\sinh (x/2)}.
\label{eq14}
\end{equation}
Therefore, the self-force will have one radial component which may be  written as
\begin{equation}
F_{\rho_{0}}=-\frac{d}{d\rho_{0}}U(\rho_{0})=\frac{q^{2}}{4\pi \rho^{2}_{0}}\Lambda(\alpha,E/E_{p}).
\label{eq14b}
\end{equation}
This result gives us a repulsive force between  the charged particle and the cosmic string in gravity's rainbow whose value depends on the gravity's rainbow function as well as on the relation between the energy of the test particle and the Planck energy.
The potential $S$ behaves as an external potential and thus causes a force on the charged particle in the space-time of a cosmic string with the deformed dispersion relation which arises from the scenario where gravity's rainbow is taken into consideration. The force associated with this potential is directly related to the parameter $\Lambda(\alpha,E/E_{p})$ and consequently the rainbow functions can be used to estimate the shape of the potential $U(\rho)$. A numerical study of function $\Lambda(\alpha,E/E_{p})$  is carried out to investigate the effect of rainbow functions and the defect parameter $\alpha$ and is shown in figures bellow.

\begin{figure}[H]
\includegraphics[scale=0.587]{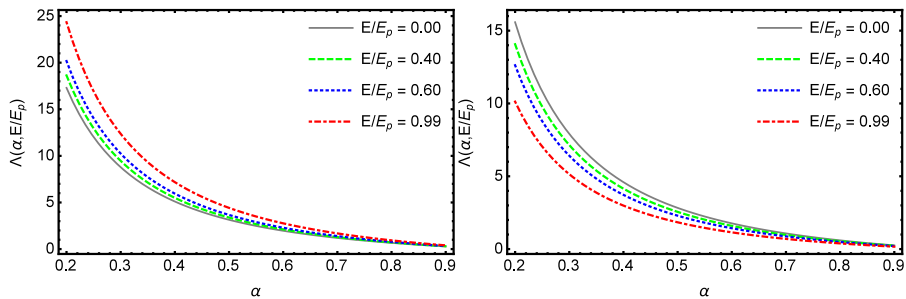}\newline
\caption{Representation of $\Lambda$ as a function of $\alpha$. On the left, $\Lambda(\alpha,E/E_{p})$ was plotted for four different values of the rainbow function defined in Eq. (\ref{eq4}), while on the right, $\Lambda(\alpha,E/E_{p})$ was plotted for values of rainbow function defined in Eq. (\ref{eq5}).}
\label{fig1}
\end{figure}

Figure \ref{fig1} shows the behavior of $\Lambda$ as a function of $\alpha$ for fixed values of $ E/E_{p}$. On the left plot, the rainbow function is given by Eq. (\ref{eq4}), while on the right side the rainbow function is defined by Eq. (\ref{eq5}). It is possible to see that the function (\ref{eq4}) increases the value of $\Lambda(\alpha,E/E_{p})$ and the rainbow function (\ref{eq5}) decreases the value of $\Lambda(\alpha,E/E_{p})$ as compared with the result obtained in the framework of the general relativity \cite{rainbow15}. Thus, the behavior of $\Lambda $ depends on the rainbow function, which implies that the intensity of the self-interaction varies according to the modifications produced, in such a way to have an induced charge which is modified not only by the deficit angle, but also by the rainbow functions. At this point, it is worth calling attention to the fact that the results of general relativity are recovered when $f_{1}(E/E_{p}) $ and $f_{2}(E/E_{p}) $ tend to 1 or, equivalently $E/E_{p}\rightarrow 0.$ This situation is represented by the solid lines in Fig.\ref{fig1}.
\section{Possible constraints on rainbow functions}
As shown in equation Eq. (\ref{eq14b}), the self-force in the space-time  of a cosmic string in gravity's rainbow has a repulsive behavior in contrast to the gravitational case where the self-force can be written as  \cite{rainbow17} 
\begin{equation}
F_{\rho_0}=-\frac{m^2 \mu}{4\pi\rho_{0}^2}\bar{\Lambda}(\alpha,E/E_{p},\mu),
\end{equation}
where $\bar{\Lambda}(\alpha,E/E_{p},\mu) =\Lambda (\alpha,E/E_{p})/\mu$ is a positive constant and $m$ is the mass of the particle. In the case of the space-time of a cosmic string in general relativity the result is obtained by taking the limit where $f_1=f_2=1$. 

Due to the fact that the self-force has only a radial component, the particle will perform a circular orbit with the speed 
\begin{equation}
v = \sqrt{\frac{m\mu\bar{\Lambda}(\alpha,E/E_{p},\mu)}{\rho_{0}}}.
\end{equation}
Therefore, we can estimate the orbital velocity of the body with a given mass and comparate with the corresponding result in the context of general relativity. For the theoretical point of view this can be done, but the constraints on the rainbow’s functions and on the parameters depends on the astronomical data which are not at our disposal at this moment. Similar analysis can be done in the case of the electrostatic self-force. Recently, we obtained some constraints on the gravity’s rainbow parameters based on results arising from LHC particle physics experiments and from Hubble Space Telescope STIS astrophysics measurements \cite{rainbow25}. Also constraints on the rainbow functions were obtained by imposing the universality of the logarithmic corrections to the semi-classical area law for the entropy of black holes \cite{rainbow26}. The result obtained in the context of gravity's rainbow adds factors associated with modified dispersion relation of the space-time, i.e., the rainbow functions that appear in the term $\Lambda(\alpha,E/E_{p})$ in Eq. (\ref{eq14}). As a result, the self-energy and the self-force have an intensity strongly connect to the choice of rainbow functions. For example equations (\ref{eq4}) and (\ref{eq5}) influence in different ways the force experienced by the test particle. Thus, observational results can significantly limit the use of certain functions in the present work.

On the other hand, the recent constraint that comes from the neutron star GW170817 \cite{ns1} observation was analyzed in the context of Rastall-Rainbow gravity in \cite{santos8} where the authors show that the rainbow function with $f_1(E/E_{p}) \geq 1$  is more appropriate than the case where  $f_1(E/E_{p})<1$. Note that for the values used for $E/E_p$ in the Figure 1, it is possible see that rainbow functions (\ref{eq4}) and (\ref{eq5})  satisfy $f_1(E/E_{p}) \geq 1$. 

Finally, let us summarize by saying that in which concerns the obtained results, the constraints could be established when we have to our disposal astronomical data concerning cosmic strings. In fact, we can only conclude that the functions we have used give us a very similar result as compared to the ones in general relativity, as expected. The only difference concerns to the intensity of the force. In one case, it is maximized when we use Eq. (\ref{eq4}), while when Eq. (\ref{eq5}) is used, this force is minimized. These results by themselves are not sufficient to discard one of the functions or to establish a constraint which means that for the phenomenon described, both functions are equally valid.

\section{Conclusion}
\label{sec4}
We have analyzed the  electrostatic self-energy and self-force associated to an electrically charged point particle induced by a
cosmic string in the context of gravity's rainbow. We have seen that a general solution may be written as the combination of a regular solution $S$ and irregular $V_{0}$ in the neighborhood of the charge, where $S$ plays  the role of an external potential. It is worth noticing that both potentials depend on the deficit angle as well as on the gravity's rainbow functions. We have also shown that self-energy depends on the rainbow functions as well as on the deficit angle. This quantity depends on the choice of the rainbow functions, in such a way that it can either increase or decrease as compared to the result obtained in the general relativity. On the right plot of Fig. \ref{fig1}, we see that the rainbow function (\ref{eq5}) decreases the intensity of the function  $\Lambda(\alpha,E/E_{p})$ and consequently reduces the absolute value of the self-energy and self-force as compared to the similar result obtained in general relativity. In contrast, for the rainbow function (\ref{eq4}), represented on the left plot,  to increase the self-energy and self-force in comparison with analogous results in the general relativity. When we consider the limits
\begin{equation}
    \lim_{E/E_{p}\rightarrow 0} f_{1} (E/E_{p})=1, \quad \lim_{E/E_{p} \rightarrow 0} f_{2} (E/E_{p})=1,
    \label{eq2}
\end{equation}
which corresponds to general relativity scenarios the results obtained are consistent with previous studies reported in the literature \cite{rainbow15}. Thus, the results obtained in the present work can be seen as a generalization of the earlier work \cite{rainbow15}, in the sense that the self-forces in each case are dependent of the energy of probe particle as compared to the Planck scale. In future works it may be interesting to extend the results obtained in this paper to backgrounds with different types of topological defects, as for example the global monopole space-time, in gravity's rainbow. Another possible extension is to consider solutions corresponding to cosmic string and global monopole in different modified theories, calculate the self-force and self-energy and analyze their characteristics and dependence with each modified theory of gravity considered.

\section{Acknowledgments}

LCNS would like to thank Conselho Nacional de Desenvolvimento Cient\'ifico e Tecnol\'ogico (CNPq) for partial financial support through the research Project No. 155361/2018-0.

\bibliographystyle{aipnum4-1}
\bibliography{referencias_unificadas}

\end{document}